\documentclass[aps,pre,reprint,twocolumn,showpacs,floatfix]{revtex4-1}

\usepackage{graphicx}
\usepackage{amsmath}
\usepackage{amssymb}
\usepackage{epsf}
\usepackage{xcolor}
\usepackage[LGRgreek]{mathastext}
\usepackage{sidecap}
\usepackage{subfigure}
\usepackage{bm}

\begin{document}

\title{The coherent feed-forward loop acts as an efficient information transmitting motif}

\author{Md Sorique Aziz Momin}
\email{soriqueaziz@jcbose.ac.in}
\affiliation{Department of Chemistry, Bose Institute, 93/1 A P C Road, Kolkata 700009, India}

\author{Ayan Biswas}
\email{ayanbiswas19@gmail.com}
\affiliation{Department of Chemistry, Bose Institute, 93/1 A P C Road, Kolkata 700009, India}

\author{Suman K Banik}
\email{skbanik@jcbose.ac.in}
\affiliation{Department of Chemistry, Bose Institute, 93/1 A P C Road, Kolkata 700009, India}

\begin{abstract} 
We present a theoretical formalism to study steady-state information transmission in coherent type 1 feed-forward loop motif with an additive signal integration mechanism. Our construct allows a two-step cascade to be slowly transformed into a bifurcation network via a feed-forward loop which is a prominent network motif. Using a Gaussian framework, we show that among these three network patterns, the feed-forward loop motif harnesses the maximum amount of Shannon mutual information fractions constructed between the final gene-product and each of the master and co-regulators of the target gene. We also show that this feed-forward loop motif provides a substantially lower amount of noise in target gene expression, compared with the other two network structures. Our theoretical predictions which remain invariant for a couple of parametric transformations, point out that the coherent type 1 feed-forward loop motif may qualify as a better decoder of environmental signals when compared with the other two network patterns in perspective.     
\end{abstract}

\date{\today}

\maketitle


\section{Introduction}

\subsection{Biochemical processes: a network-centric approach}

Gene transcription regulatory network (GTRN) is a key player in several physiological processes, e.g., sensory and developmental signaling programs, where it produces gene-products some of which namely the transcription factors (TFs) are involved in the regulation process itself. TFs are broadly categorized into activators and repressors, the former being involved in up-regulating the expression of its target gene while the latter down-regulate the effector gene. In this fashion, TFs act as the interaction mediators among several genes, thus forming several interconnected patterns like cascades, loops etc. The resulting networks have genes as nodes and mainly transcriptional regulation as edges though these may also involve other interactions in post-transcriptional, (post)translational levels \cite{Alon2006}. It has been recently found out in experiments supported by theoretical frameworks that this vast chunk of networks is made of simpler building blocks which appear with higher frequencies in a real network than in a random one and with which the nomenclature of `network motifs' has been associated in the literature \cite{Milo2002}. These network motifs are abundant in model systems like \textit{E. coli, S. cerevisiae}, mouse and also found in humans.

Three nodes interacting with each other generate an ensemble of thirteen distinct patterns or subgraphs of which only one namely the feed-forward loop (FFL) has been observed to function as a network motif in bacterial GTRN as well as in higher organisms \cite{Alon2006}. In an FFL, a TF (S) regulates another TF (X) whereas both of them regulate the target gene (Y). Hence, often S and X are termed as the master and co-regulator of Y, respectively. Depending upon the combinations of activation/repression for the three edges, there arise a total of eight types of FFLs, divided into two major sub-groups. Designating $+~(-)$ for activating (repressing) edges, the effective signature of the indirect regulation of Y by S (via X) is determined by the product of signatures of two consecutive regulatory edges (i.e., from S to X and from X to Y). If the effective signature of the indirect pathway matches with that of the direct one (from S to Y), FFL is of coherent type and incoherent otherwise. Among these eight types, only two are abundant in nature, namely coherent (C) and incoherent (I) type 1 FFL, the former being the model motif of our present analysis \cite{Mangan2003a}. C1 FFL comprises of three activating edges and is utilized in arabinose and flagella system of \textit{E. coli} where the input functions similar to AND gate and OR gate are utilized, respectively to integrate activities of upstream TFs (S, X) in the target gene (Y). The former is found to implement a sign-sensitive delay for ON steps \cite{Mangan2003b}, while the latter introduces the delay for OFF steps \cite{Kalir2004} of the signal corresponding to S.


\begin{figure}[!h]
\includegraphics[width=0.9\columnwidth,angle=0]{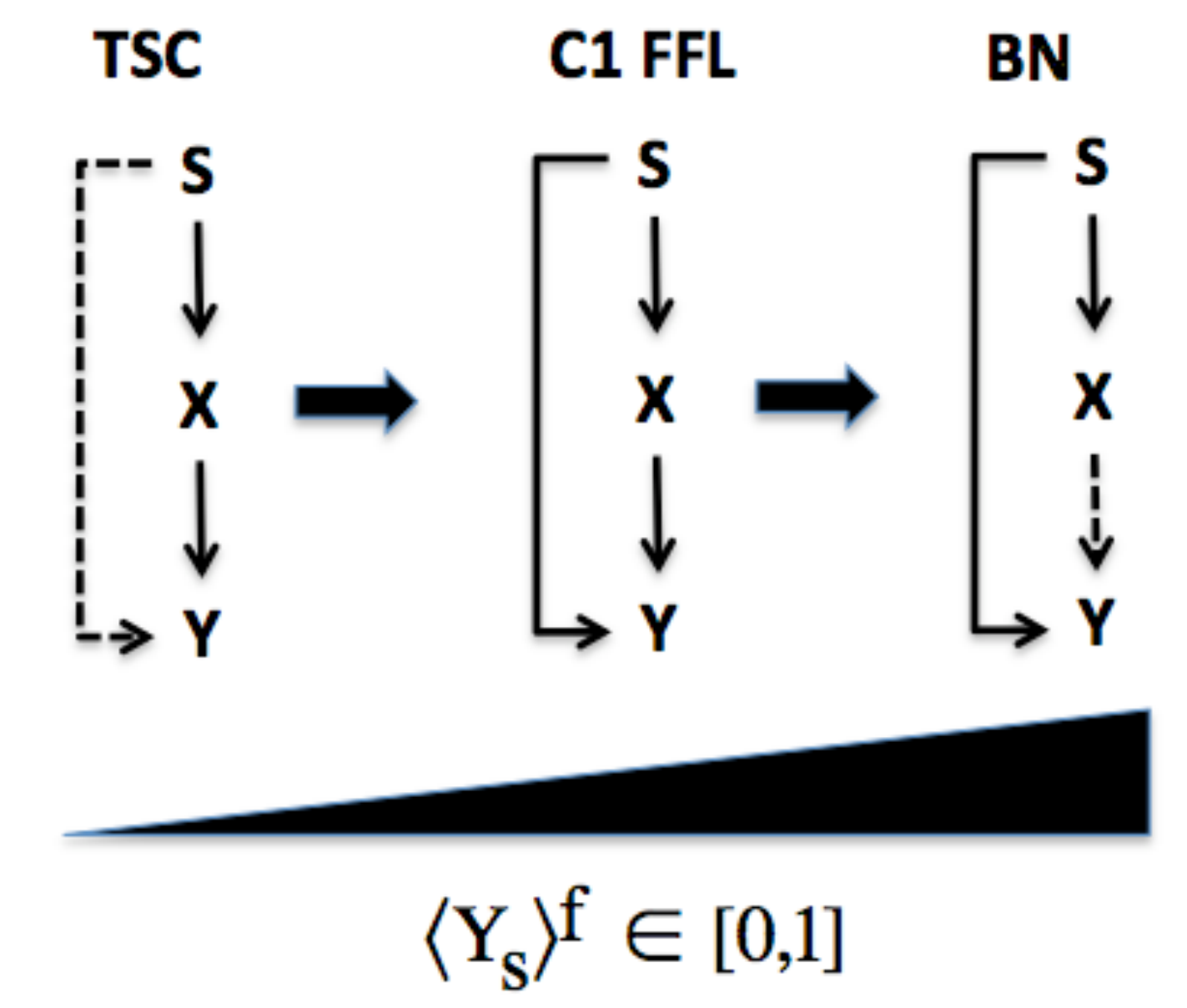} 
\caption{Schematic diagram depicting the theoretical construct which gradually enables S mediated synthesis of Y ($S \rightarrow Y$ producing $\langle \text{Y}_{\text{s}} \rangle$ copies) and consequent disabling of X mediated synthesis of Y ($X \rightarrow Y$ producing $\langle \text{Y}_{\text{x}} \rangle$ copies) so that steady-state population levels of S, X and, Y ($\langle \text{Y} \rangle=\langle \text{Y}_{\text{s}} \rangle+\langle \text{Y}_{\text{x}} \rangle$ copies) remain fixed at each step of topological modifications (Thick arrows). Hence, a TSC is modified into a C1 FFL and finally into a BN. Thin solid (dashed) arrows symbolize the presence (absence) of the corresponding genetic interactions. $\langle \text{Y}_{\text{s}} \rangle^\text{f} =: \langle \text{Y}_{\text{s}} \rangle/\langle \text{Y} \rangle$ acts as the normalized tuning parameter in our analysis.}
\label{fig1}
\end{figure}

Our present framework is designed to evaluate the effect of topological features of a C1 FFL on its information processing capabilities at steady state. For that purpose, we start from a network pattern where the direct path $S \rightarrow Y$ is absent, and Y is produced entirely from contributions of X via $X \rightarrow Y$. We gradually empower the former edge (contributing $\langle Y_s \rangle$ copies) and correspondingly weaken the latter (contributing $\langle Y_x \rangle$ copies) so that at each of the resulting network configurations, steady-state population levels of Y ($\langle Y \rangle = \langle Y_s \rangle + \langle Y_x \rangle$) remain fixed. Here, $\langle ... \rangle$ denotes steady-state ensemble average. The edge $S \rightarrow X$ is maintained unaltered with fixed steady-state populations of the respective gene-products. The two extreme network conformations that we get from our construct are a two-step cascade (TSC), and a bifurcation network (BN) whereas all the intermediate conformations, in principle, fall into the group of C1 FFL (see schematic Fig.\ref{fig1}). It is to be noted that neither TSC nor BN is a network motif whereas C1 FFL is a network motif with the highest relative abundance among the total eight types of FFL found in \textit{E. coli} and \textit{S. cerevisiae}. We modeled this genetic motif conforming to a Gaussian framework where gene-products S, X, and Y undergo production in and removal (degradation and/or dilution) from a unit effective cellular volume, described neatly in a Langevin formalism.

\subsection{Information theory applied to biological systems}

The inherent stochastic behavior of the biochemical species can be quantified using the concept of entropy of the associated random variables. In other words, the entropy of a gene-product quantifies the extent of its uncertainty or fluctuations. The inter-specific coupling of two gene-products makes their individual (marginal) entropy spaces to overlap with each other, the common or shared region denoting the mutual information (MI) between them. MI is symmetric with respect to its argument variables. MI, being a nonlinear metric of association or correlation denotes in our present case, how much uncertainty about a gene-product can be reduced on average if knowledge of another gene-product is available and vice versa \cite{Cover1991,MacKay2002}. This concept is readily generalizable to include three variables where one quantifies how much fluctuation space a particular random variable shares with a group of two other random variables. For a pictorial depiction of MI, we refer our readers to the Venn diagrams in Appendix Figs.~(\ref{figa1}-\ref{figa3}).

Information theory established by Claude Shannon \cite{Shannon1948}, has been used in recent years to deal with signal processing in biological systems primarily in the perspective of neuroscience  \cite{Borst1999,Timme2014,Timme2018}. One particular information-theoretic approach to understand the biological phenomenon is Ref.~\cite{Mehta2009} where MI between input signals and output response in \textit{V. harveyi} quorum-sensing network was quantified. The network structure used by this model of the marine bacterium to process incoming autoinducer signals is of a converging (integrating) type similar to the FFL motif. Applications to other biochemical networks also made significant contributions making accurate predictions regarding developmental processes in fruit fly embryo \cite{Bialek2012}; others linked network topology with information processing capabilities \cite{Cheong2011,Tareen2018} whereas research was also directed to estimate the effect of various noise sources in a cellular population on its transmitted information content \cite{Chevalier2015}. Information theory has also been used interfacing with experiments on yeast TF Msn2 to find high fidelity information transduction of signal identity in contrast with signal intensity \cite{Hansen2015}. Previous studies also shed light on information-theoretic connotations of biological fitness \cite{Taylor2007,Tareen2018}. Another important study focussed on information optimization in a genetic non-loop feed-forward pattern \cite{Walczak2010}. In Ref.~\cite{Ronde2012}, the FFL has been shown to act as filters when exposed to temporally varying signals.

In this article, we envisage formulating an information-theoretic description of steady-state signal processing comparing TSC, C1 FFL, and BN. In the following \textit{Model} section, we derive key statistical elements, necessary to build up MI-based understanding of signal processing in these three network structures. Pathway-specific contributions in the fluctuations of the target gene-expression are also computed in closed-form analytical expressions. In the \textit{Results and Discussion} section, we provide the reason for proposing new metrics beyond the conventional Shannon MI. The predictions emanating from these derived quantifiers of stochastic signal processing are justified by another metric measuring the copy number variability of the output gene-product. This latter metric has been tailor-made for our present construct of network-architecture modification, keeping in perspective that biological noise is primarily generated to randomized production and degradation events of various biochemical species. These theoretical predictions are also shown to possess interesting invariant features corresponding to some specific and biologically important parametric transformations. Finally, we draw \textit{Conclusions} summarizing our key theoretical predictions and putting forward comments on future research directions.


\section{The Model}

\subsection{Langevin description of gene-expression in a typical FFL}

In a C1 FFL, S activates X and both of them activate Y using an additive synthesis mechanism and the resulting dynamics of the copy numbers of S, X, and Y in a unit effective cellular volume can be represented using Langevin formalism as \cite{Biswas2016,Biswas2018,Biswas2019}, 
\begin{subequations}
\begin{eqnarray}
\label{eq1a}
\frac{dS}{dt} & = & f_s - \mu_s S + \xi_s(t), \\
\label{eq1b}
\frac{dX}{dt} & = & f_x (S) - \mu_x X + \xi_x(t), \\
\label{eq1c}
\frac{dY}{dt} & = & f_y (S,X) - \mu_y Y + \xi_y(t).
\end{eqnarray}
\end{subequations}

\noindent The synthesis rates are $f_{s} = k_{s}~\text{for}~\phi \rightarrow S$, $f_{x} = k_{sx} (S^n/(S^n+K_{sx}^n))~\text{for}~S \rightarrow X$, $f_{y} = k_{sy}(S^n/(S^n+K_{sy}^n)) + k_{xy}(X^n/(X^n+K_{xy}^n))~\text{for}~S \rightarrow Y~Add~X \rightarrow Y$ \cite{Bintu2005b}. k and $\mu$ are the synthesis and degradation rate parameters whereas the Hill function $(S^n/(S^n+K_{sx}^n))$ denotes the occupancy probability of promoter in gene X by S with cooperativity index `n' and activation coefficient $K_{sx}$ for the transcriptional event $S \rightarrow X$ and so on. Noise terms ($\xi_i, i = s,~x,~and~y$) are modeled as independent white Gaussian types with $\langle \xi_i (t) \rangle = 0$ and $\langle \xi_i (t) \xi_j (t') \rangle =  \langle | \xi_i |^2 \rangle \delta_{ij} \delta (t-t')$ \cite{Tostevin2010}. $\langle...\rangle$ is used to denote steady-state ensemble average. Noise strengths are generated equally from the synthesis and degradation of the biochemical species at steady state. For species S, it is expressed as $\langle | \xi_s |^2 \rangle = \langle f_s \rangle+\mu_s \langle S \rangle = 2\langle f_s \rangle~(= 2\mu_s \langle S \rangle)$, which is an approximation. Steady-state noise strengths for X and Y are modeled similarly. Taking recourse to the linear noise approximation (LNA) \cite{Keizer1987,Elf2003,Kampen2007} and Lyapunov equation at steady state \cite{Paulsson2004,Paulsson2005}, we arrive at the set of closed-form analytic expressions for the second moments associated with Gaussian random variables representing the three gene-products. Since the distributions of environmental stimuli, which an organism encounters in its natural setting, are not experimentally measured in most of the cases, the Gaussian assumption is preferred as it gives the lower bounds of the various MI \cite{Mitra2001} computed considering the gene-products.

\subsection{Analytical expressions of the second moments}

We linearize the set of Eqs.~1(a-c) at steady state, by taking into account small perturbations of the kind $\delta Z(t) = Z(t)-\langle Z \rangle \ll \langle Z \rangle $. Z symbolizes the random variable corresponding to the set of copy numbers \{S, X, Y\} of associated gene-products. Invoking the steady-state Lyapunov equation to connect fluctuations and dissipation in our present construct, we are able to extract the second moments, i.e., variances and covariances of the gene-products.
\begin{equation}
\label{eq2}
\bm{J}\bm{\Sigma}+\bm{\Sigma}\bm{J}^T+\bm{D}=\bm{0},
\end{equation}  
with
\begin{eqnarray*}
\bm{J} = \left (
\begin{array}{ccc}
 - \mu_s & 0 & 0 \\
\langle f^{\prime}_{x,s} \rangle & 
 - \mu_x & 0 \\
\langle f^{\prime}_{y,s} \rangle & 
\langle f^{\prime}_{y,x} \rangle & 
 - \mu_y
\end{array} 
\right ),
\bm{\Sigma} = \left (
\begin{array}{ccc}
 \Sigma(S) & \Sigma(S,X) & \Sigma(S,Y) \\
\Sigma(X,S) & 
 \Sigma(X) & \Sigma(X,Y) \\
\Sigma(Y,S) & 
\Sigma(Y,X) & 
\Sigma(Y)
\end{array} 
\right ),
\nonumber \\
\; \text{and} \;
\mathbf{D} = \left (
\begin{array}{ccc}
 \langle |\xi_{s}|^2 \rangle & 0 & 0 \\
0 & 
  \langle |\xi_{x}|^2 \rangle & 0 \\
0 & 
0 & 
 \langle |\xi_{y}|^2 \rangle
\end{array} 
\right ).
\end{eqnarray*}\\
Jacobian $\bm{J}$ quantifies how the deterministic parts of the Langevin equations respond to small steady-state changes in the random variables (copy numbers) whereas $\bm{\Sigma}$ contains (co)variances of them. We use $\langle f^{\prime}_{x,s} \rangle =: \left ( \frac{df_x(S)}{dS} \right )_{S = \langle S \rangle}$ and so on. Due to the fact that $\Sigma(X,S) = \Sigma(S,X)$ etc., $\bm{\Sigma}$ becomes a symmetric matrix. $T$ denotes matrix transposition operation. $\bm{D}$ contains various steady-state noise strengths and due to the absence in noise correlation between different biochemical species, it is diagonal in construction. Solving Eq.(\ref{eq2}) for $\bm{\Sigma}$, the following expressions are obtained. 
\begin{subequations}
\begin{eqnarray}
\label{eq3a}
\Sigma(S) & = & \frac{\langle |\xi_{s}|^2 \rangle}{2\mu_{s}}, \nonumber \\
& = & \langle S \rangle.\\
\label{eq3b}
\Sigma(S,X) & = &  \frac{\langle f_{x,s}' \rangle \Sigma (S)}{(\mu_{s}+\mu_{x})}, \nonumber \\
& = & \left ( \frac{nK^n_{sx}}{{\langle S \rangle}^n+K^n_{sx}} \right ) \frac{\langle X \rangle}{\langle S \rangle} \Phi_{xs} \Sigma(S). \\ 
\label{eq3c}
\Sigma(S,Y)  & = &  \frac{\langle f_{y,s}' \rangle \Sigma (S) + \langle f_{y,x}' \rangle \Sigma (S,X)}{(\mu_{s}+\mu_{y})}, \nonumber \\
& = & \left ( \frac{nK^n_{sy}}{{\langle S \rangle}^n+K^n_{sy}} \right ) \frac{\langle Y \rangle}{\langle S \rangle} {\langle Y_s \rangle}^f \Phi_{ys} \Sigma(S) \nonumber \\
& + & \left ( \frac{nK^n_{xy}}{{\langle X \rangle}^n+K^n_{xy}} \right ) \frac{\langle Y \rangle}{\langle X \rangle} {\langle Y_x \rangle}^f \Phi_{ys} \Sigma(S,X). \\ 
\label{eq3d}
\Sigma(X) & = &  \frac{\langle |\xi_{x}|^2 \rangle}{2\mu_{x}} + \frac{\langle f_{x,s}' \rangle \Sigma (S,X)}{\mu_{x}}, \nonumber \\
& = & \langle X \rangle + \left ( \frac{nK^n_{sx}}{{\langle S \rangle}^n+K^n_{sx}} \right ) \frac{\langle X \rangle}{\langle S \rangle} \Sigma(S,X). \\
\label{eq3e}
\Sigma(X,Y)  & = &  \frac{\langle f_{y,s}' \rangle \Sigma (S,X) + \langle f_{x,s}' \rangle \Sigma (S,Y)+\langle f_{y,x}' \rangle \Sigma (X)}{(\mu_{x}+\mu_{y})}, \nonumber \\
& = & \left ( \frac{nK^n_{sy}}{{\langle S \rangle}^n+K^n_{sy}} \right ) \frac{\langle Y \rangle}{\langle S \rangle} {\langle Y_s \rangle}^f \Phi_{yx} \Sigma(S,X) \nonumber \\
& + & \left ( \frac{nK^n_{sx}}{{\langle S \rangle}^n+K^n_{sx}} \right ) \frac{\langle X \rangle}{\langle S \rangle} \Phi_{xy} \Sigma(S,Y) \nonumber \\
& + & \left ( \frac{nK^n_{xy}}{{\langle X \rangle}^n+K^n_{xy}} \right ) \frac{\langle Y \rangle}{\langle X \rangle} {\langle Y_x \rangle}^f \Phi_{yx} \Sigma(X).
\end{eqnarray}
\begin{eqnarray}
\label{eq3f}
\Sigma(Y)  & = &  \frac{\langle |\xi_{y}|^2 \rangle}{2\mu_{y}} + \frac{\langle f_{y,s}' \rangle \Sigma (S,Y) + \langle f_{y,x}' \rangle \Sigma (X,Y)}{\mu_{y}}, \nonumber \\
& = & \langle Y \rangle + \left ( \frac{nK^n_{sy}}{{\langle S \rangle}^n+K^n_{sy}} \right ) \frac{\langle Y \rangle}{\langle S \rangle} {\langle Y_s \rangle}^f \Sigma(S,Y) \nonumber \\
& + & \left ( \frac{nK^n_{xy}}{{\langle X \rangle}^n+K^n_{xy}} \right ) \frac{\langle Y \rangle}{\langle X \rangle} {\langle Y_x \rangle}^f \Sigma(X,Y).
\end{eqnarray} 
\end{subequations}

\begin{widetext}

\noindent Here, $\Phi_{ij} =: \frac{\mu_i}{\mu_i + \mu_j}~\text{with}~i,j(\neq i) \in \{s,x,y\}$ and is  referred as `filtering function' since $\mu_i = {\tau_i}^{-1}$, $\tau_{i}$ being the average life-time of species `i' and therefore $\Phi_{ij}$ quantifies the separation of relaxation time scales between any two species `i' and `j', controlling flow of fluctuations downstream in the motif. From these expressions, one can also evaluate partial variances in the following manner \cite{Barrett2015}.
\begin{subequations}
\begin{eqnarray}
\label{eq4a}
\Sigma(S|Y) & = & \Sigma(S)-\frac{\Sigma^2(S,Y)}{\Sigma(Y)}, \\
\label{eq4b}
\Sigma(S|X,Y) & = & \Sigma(S)-\frac{\Sigma(X)\Sigma^2(S,Y)-2\Sigma(S,X)\Sigma(S,Y)\Sigma(X,Y)+\Sigma(Y)\Sigma^2(S,X)}{\Sigma(X)\Sigma(Y)-\Sigma^2(X,Y)}.
\end{eqnarray}
\end{subequations} 
In a similar fashion, one can write expressions for $\Sigma(X|Y)~and~\Sigma(X|Y,S)$. These reductions in variances of a single designated species while conditioning upon one or two other species are necessary ingredients to compute MI terms. For Gaussian random variables, MI is expressible as a logarithmic functional of second moments, e.g., MI constituted between S and Y is $I(S;Y)=(1/2)\log_{2}(\Sigma(S)/\Sigma(S|Y))$ in the units of `bits'. Analogously, three-variable MI between S and (X,Y) is expressed as $I(S;X,Y)=(1/2)\log_{2}(\Sigma(S)/\Sigma(S|X,Y))$ \cite{Barrett2015}. Similarly, one can arrive at other MI, e.g., I(X;Y), I(X;Y,S).

\subsection{Decomposition of variance of target gene-product}

Starting from Eq.(\ref{eq3f}), $\Sigma(Y)$ can be written explicitly in terms of systemic parameters along with $\langle S \rangle$, $\langle X \rangle$, $\langle Y \rangle$, ${\langle Y_s \rangle}^f$ etc. to characterize contributions of a one-step cascade or OSC (S $\rightarrow$ Y), a TSC (S $\rightarrow$ X $\rightarrow$ Y) and a cross term (CT).
\begin{subequations}
\begin{eqnarray}
\label{eq5a} 
\Sigma(Y) & = & \Sigma(Y)_{OSC} + \Sigma(Y)_{TSC} + \Sigma(Y)_{CT}. \\
\label{eq5b}
\Sigma(Y)_{OSC} & = & \langle Y \rangle {\langle Y_s \rangle}^f + \left ( \frac{nK^n_{sy}}{{\langle S \rangle}^n+K^n_{sy}} \right )^2\frac{{\langle Y \rangle}^2}{\langle S \rangle}{{\langle Y_s \rangle}^f}^2 \Phi_{ys}, \nonumber \\
& = & c_1 {\langle Y_s \rangle}^f + c_2 {{\langle Y_s \rangle}^f}^2. \\
\label{eq5c}
\Sigma(Y)_{TSC} & = & \langle Y \rangle {\langle Y_x \rangle}^f + \left ( \frac{nK^n_{xy}}{{\langle X \rangle}^n+K^n_{xy}} \right )^2 \frac{{\langle Y \rangle}^2}{\langle X \rangle} {{\langle Y_x \rangle}^f}^2 \Phi_{yx} \nonumber  \\
& + & \left ( \frac{nK^n_{sx}}{{\langle S \rangle}^n+K^n_{sx}} \right )^2\left ( \frac{nK^n_{xy}}{{\langle X \rangle}^n+K^n_{xy}} \right )^2 \frac{{\langle Y \rangle}^2}{\langle S \rangle} {{\langle Y_x \rangle}^f}^2 (\Phi_{yx}\Phi_{xs}+\Phi_{ys}\Phi_{xs}\Phi_{xy}), \nonumber \\
& = & c_3 {\langle Y_x \rangle}^f + c_4 {{\langle Y_x \rangle}^f}^2 = c_3 (1-{\langle Y_s \rangle}^f) + c_4 {(1-{\langle Y_s \rangle}^f)}^2. \\
\label{eq5d} 
\Sigma(Y)_{\text{CT}} & = & \left ( \frac{nK^n_{sy}}{{\langle S \rangle}^n+K^n_{sy}} \right )\left ( \frac{nK^n_{xy}}{{\langle X \rangle}^n+K^n_{xy}} \right )\left ( \frac{nK^n_{sx}}{{\langle S \rangle}^n+K^n_{sx}} \right )\frac{{\langle Y \rangle}^2}{\langle S \rangle} {\langle Y_s \rangle}^f {\langle Y_x \rangle}^f (\Phi_{ys}\Phi_{xs}+\Phi_{ys}\Phi_{xy}+\Phi_{yx}\Phi_{xs}), \nonumber \\ 
& = & c_5 {\langle Y_s \rangle}^f {\langle Y_x \rangle}^f = c_5 {\langle Y_s \rangle}^f (1-{\langle Y_s \rangle}^f). 
\end{eqnarray}
\end{subequations}
\noindent $\text{Where,}~
c_1 =:  \langle Y \rangle,~
c_2 =: \left ( \frac{nK^n_{sy}}{{\langle S \rangle}^n+K^n_{sy}} \right )^2\frac{{\langle Y \rangle}^2}{\langle S \rangle}\Phi_{ys},~
c_3 =: \langle Y \rangle,\\
c_4 =: \left ( \frac{nK^n_{xy}}{{\langle X \rangle}^n+K^n_{xy}} \right )^2\frac{{\langle Y \rangle}^2}{\langle X \rangle}\Phi_{yx}+\left ( \frac{nK^n_{sx}}{{\langle S \rangle}^n+K^n_{sx}} \right )^2\left ( \frac{nK^n_{xy}}{{\langle X \rangle}^n+K^n_{xy}} \right )^2\frac{{\langle Y \rangle}^2}{\langle S \rangle}(\Phi_{yx}\Phi_{xs}+\Phi_{ys}\Phi_{xs}\Phi_{xy}), \\
c_5 =: \left ( \frac{nK^n_{sy}}{{\langle S \rangle}^n+K^n_{sy}} \right )\left ( \frac{nK^n_{xy}}{{\langle X \rangle}^n+K^n_{xy}} \right )\left ( \frac{nK^n_{sx}}{{\langle S \rangle}^n+K^n_{sx}} \right )\frac{{\langle Y \rangle}^2}{\langle S \rangle}(\Phi_{ys}\Phi_{xs}+\Phi_{ys}\Phi_{xy}+\Phi_{yx}\Phi_{xs}).
$

\end{widetext}


\section{Results and Discussion}

\subsection{Roles of different gene-products in distinct network patterns}

We increase $k_{sy}$ so as to increase $\langle Y_s \rangle$ by 1 copy at an instant and also decrease $k_{xy}$ and consequently $\langle Y_x \rangle$ by 1 copy so that $\langle Y \rangle$ remains fixed at 100 copies in a unit effective cellular volume. Thus, the direct transcriptional edge $S \rightarrow Y$ is gradually strengthened and indirect transcriptional edge $X \rightarrow Y$ is equally weakened. The synthesis rate parameters, $k_{sy}$ and $k_{xy}$, are the maximal expression levels of the target gene Y, regulated by S and X, respectively. Changing these rate parameters results in controlling the RNA polymerase as a result of which mRNA production per unit time is changed \cite{Alon2006}. $\langle S \rangle = 10, \langle X \rangle = 100~\text{copies}$ are also maintained throughout the entire process. Keeping gene-expression levels fixed while the three-node topology in perspective is discretely modified, helps to compare different network architectures on an equal footing \cite{Alon2006}. Whenever the construct allows signal S and X to converge onto Y, this takes recourse to an additive integration mechanism. We start with $k_{sy} = 0$ ($k_{xy}$ maximum) that makes $\langle Y_s \rangle = 0$ ($\langle Y_x \rangle = \langle Y \rangle$) and the resulting network is a TSC, $S \rightarrow$ X $\rightarrow Y$. Here, the signal from the input TF (S) is relayed to the final gene-product (Y) via the intermediate TF (X). TSC is typically found in transcriptional networks engaged in slow and irreversible processes like developmental phenomena in \textit{sea urchin}, \textit{D. melanogaster} etc \cite{Rosenfeld2003}. The other end where $k_{sy}$ is maximum ($k_{xy} = 0$) i.e., $\langle Y_s \rangle = \langle Y \rangle$ ($\langle Y_x \rangle = 0 $) is a BN. In a BN, S activates two of its downstream gene-products X and Y which do not have any direct interaction between them. BN is composed of two one-step cascade (OSC) namely, S $\rightarrow$ X and S $\rightarrow$ Y. OSC is important for sensory signal transduction which is supposed to be fast and reversible \cite{Rosenfeld2003}. All the intermediate operating points for which $\langle Y_s \rangle$ ($\langle Y_x \rangle) = 1 - 99$ copies, designate to C1 FFL where S activates the synthesis of X and both of them activates the production of the final gene-product Y. In other words, both the direct and indirect regulatory branches contribute to the production of Y, which always maintains a constant pool of 100 copies in a unit effective cellular volume. The entire range of $\langle Y_s \rangle^f$ ($\langle \text{Y}_{\text{s}} \rangle^\text{f} =: \langle \text{Y}_{\text{s}} \rangle/\langle \text{Y} \rangle$) keeps genetic identities of S and Y (a regulatory input and a target gene-product, respectively) same throughout but not for X which is an intermediate TF in both TSC and FFL but in BN it is one of the effector gene-products, another being Y. Therefore in TSC, X and Y are the encoded and decoded signals, respectively. FFL also treats X as an encoded signal whereas Y is the culmination of direct decoding (through $\langle Y_s \rangle$) and indirect decoding (through $\langle Y_x \rangle$) of input S. In a BN, both X and Y cherish the status of the direct decoded signal via two OSCs. The correlation developed between X and Y in a BN is thus completely assisted by the fluctuation space of S. TSC has been found to operationalize synergistic and redundant information in its embedded information channels \cite{Biswas2016,Biswas2019}. On the other hand, BN may work as an upper-level sub-motif creating information redundancy in a diamond motif which is a variant of FFL with duplication of intermediate gene X \cite{Biswas2018}.

\subsection{Two and three-variable MI and fluctuations in target gene-expression}

The inter-genetic interactions among S, X, and Y being essentially nonlinear augmented by cooperative binding of TF molecules to their target promoter sites, the correlation structure is suitably quantified using MI rather than simple linear measures, e.g., Pearson correlation coefficient. Taking into consideration all the three different types of network manifestations coming out of our theoretical construct, it can be generally stated that I(S;Y) is the overlap between entropy spaces of the first-level signaling TF S and the utmost downstream (one of the two in the case of BN) gene-product Y. Similarly, I(X;Y) is the shared proportion of entropic contributions of two downstream gene-products X (intermediate TF in TSC and FFL) and Y (effector gene-product). I(S;X,Y) is I(S;Y) added with MI of S and X given the knowledge of Y, i.e., I(S;X$|$Y). Similarly, adding MI between X and S given the knowledge about Y, i.e., I(X;S$|$Y) (= I(S;X$|$Y)) to I(X;Y) amounts to a production of I(X;Y,S). These MI terms are related to each other forming chain rules  \cite{Cover1991,MacKay2002} 
\begin{subequations}
\begin{eqnarray}
\label{eq6a}
I(S;X,Y) & = & I(S;Y)+I(S;X|Y), \\
\label{eq6b}
I(X;Y,S) & = & I(X;Y)+I(X;S|Y).
\end{eqnarray}
\end{subequations}


\begin{figure*}[!t]
\includegraphics[width=1.8\columnwidth,angle=0]{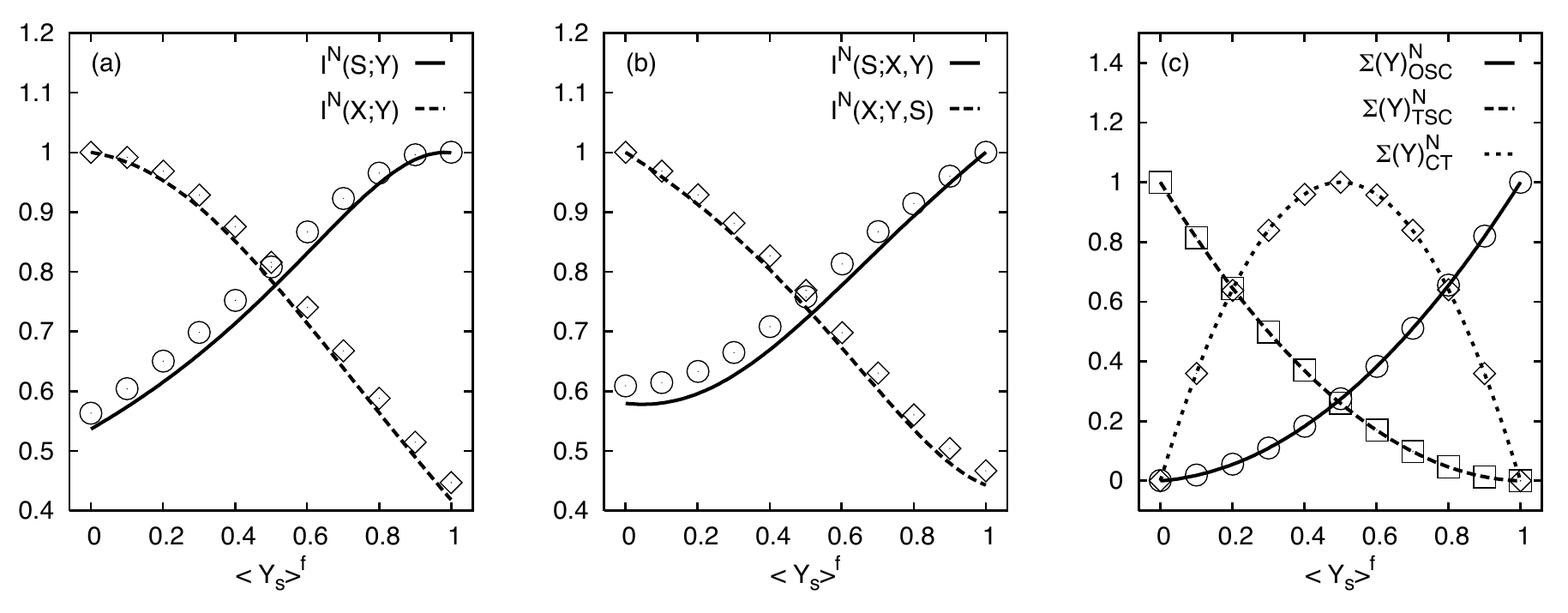}
\caption{(a) Two-variable MI, I(S;Y) and I(X;Y) normalized by their respective maximum value, producing I$^N$(S;Y) and I$^N$(X;Y), respectively. Similarly, (b) normalized three-variable MI, I$^N$(S;X,Y) and I$^N$(X;Y,S), and (c) normalized constituents of $\Sigma(Y)$ i.e., $\Sigma(Y)^N_{OSC},~\Sigma(Y)^N_{TSC},~and~\Sigma(Y)^N_{CT}$. The horizontal axis represents $\langle Y_s \rangle^f$. As the edge $S \rightarrow Y~(X \rightarrow Y)$ is gradually enabled (disabled) as a result of which, $\langle Y_s \rangle^f (\langle Y_x \rangle^f$) increases (decreases); I$^N$(S;Y) and I$^N$(S;X,Y) (I$^N$(X;Y) and I$^N$(X;Y,S)) increases (decreases). One striking feature is the cross over regime between I$^N$(S;Y) and I$^N$(X;Y) and also between I$^N$(S;X,Y) and I$^N$(X;Y,S), occurring when strengths of both $S \rightarrow Y ~and~ X \rightarrow Y$ are approximately equal and therefore $\langle Y_s \rangle^f \approx 0.5 \approx \langle Y_x \rangle^f$.  $\Sigma(Y)^N_{OSC}$ and $\Sigma(Y)^N_{TSC}$ intersect each other and the peak of $\Sigma(Y)^N_{CT}$ appears exactly at $\langle Y_s \rangle^f = 0.5=\langle Y_x \rangle^f $. Population levels are fixed at $\langle S \rangle = 10,~\langle X \rangle = 100,~and~\langle Y \rangle = 100 ~ \text{copies per unit effective cellular volume}$. Relaxation rate parameters are $\mu_{s} = 0.1,~\mu_{x} = 0.5,~and~\mu_{y} = 5.0 ~\text{all in the units of}~min^{-1}$. Synthesis rate parameters are determined according to $k_s=\mu_{s}\langle S \rangle,~k_{sx}=\mu_x\langle X \rangle(\langle S \rangle^n/(\langle S \rangle^n+K_{sx}^n))^{-1},~k_{sy}=\mu_y\langle Y_s \rangle(\langle S \rangle^n/(\langle S \rangle^n+K_{sy}^n))^{-1},~\text{and}~k_{xy}=\mu_y\langle Y_x \rangle(\langle X \rangle^n/(\langle X \rangle^n+K_{xy}^n))^{-1}$. Here, Hill coefficient $n = 2$ and activation coefficients are $K_{sx} = \langle S \rangle,~K_{sy} = \langle S \rangle,~and~K_{xy} = 2\langle X \rangle$. These parameters effectively make the occupancy probability of promoter Y by X, 0.2 which is characteristic of linear activation whereas other regulatory edges employ half-maximal activation. Lines are drawn using analytical results obtained from LNA. These are supported by \textit{in silico} results obtained by averaging over an ensemble of 10$^5$ independent samples derived from Gillespie SSA \cite{Gillespie1976,Gillespie1977} and are drawn as symbols. }
\label{fig2}
\end{figure*}

Fig.~\ref{fig2}(a-b) includes the profiles of several MI normalized by their respective maximum value (denoted by superscript `N'), thereby showing their nature of variations subject to changing $\langle Y_s \rangle^f$. As the contributions of direct path (indirect path) increases (decreases), $I^N(S;Y)$ ($I^N(X;Y)$) grows (decays) in Fig.~\ref{fig2}(a). These trends are also manifested for normalized three-variable MI, $I^N(S;X,Y)$ and $I^N(X;Y,S)$ in Fig.~\ref{fig2}(b). Since the target gene-product is generally involved in a number of key physiological processes in the organism, the fluctuation level associated with the corresponding gene-expression qualifies to be an important quantifier of systemic performance. In this regard, we showcase Fig.~\ref{fig2}(c) containing profiles of normalized components pertaining to contributions from OSC, TSC, and CT in $\Sigma(Y)$. Consulting Eqs.(\ref{eq5a}-\ref{eq5d}), the trends of these components can be rationalized at ease. While $\Sigma(Y)^N_{OSC}$ has a growing nature, $\Sigma(Y)^N_{TSC}$ decays whereas the profile of $\Sigma(Y)^N_{CT}$ is concave down.

\subsection{MI fraction: a better quantifier of information processing}

The individual MI terms do not distinguish FFL and the same can be said for $\Sigma(Y)^N_{OSC}$ and $\Sigma(Y)^N_{TSC}$. $\Sigma(Y)^N_{CT}$ is maximum where $\langle Y_s \rangle^f = 0.5$ i.e., both the direct and the indirect branch of FFL contributes equally. These observations lead us to search for suitable multivariate information-theoretic metrics that can provide insights into the signal processing function of a typical C1 FFL topology. To this end, we define two rescaled metrics
\begin{subequations}
\begin{eqnarray}
\label{eq7a}
I^f (S;Y) & =: & I(S;Y)/I(S;X,Y), \\
\label{eq7b}
I^f (X;Y) & =: & I(X;Y)/I(X;Y,S).     
\end{eqnarray}
\end{subequations}


\begin{figure*}[!t]
\includegraphics[width=1.8\columnwidth,angle=0]{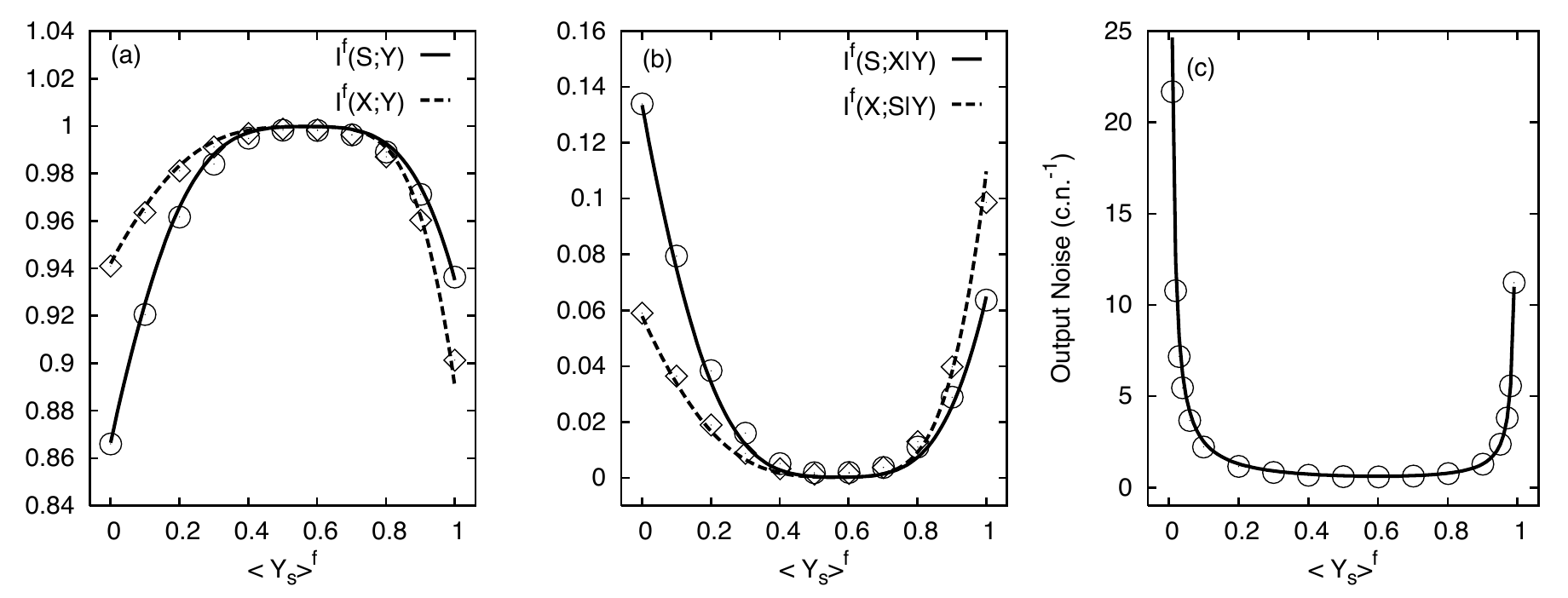}
\caption{(a) I$^f$(S;Y), I$^f$(X;Y); (b) I$^f$(S;X$|$Y), I$^f$(X;S$|$Y); (c) Output Noise (in units of ${c.n.}^{-1}$), all with respect to $\langle Y_s \rangle^f$. In panel (a), we find that parameters belonging to a C1 FFL provide the maximum amount of I$^f$(S;Y), I$^f$(X;Y) ($\approx 1$) than the TSC and BN, thereby forming a plateau. In panel (b), partial MI fractions show antagonistic nature to profiles of previous MI fractions in panel (a). Panel (c) shows that C1 FFL delivers minimal Output Noise, supporting maximal MI fractions achievable by the motif, as depicted in panel (a). Here, analytical results are shown with lines. Accompanying simulation data (average of 10$^5$ independent time series) using Gillespie SSA are denoted by symbols. Parameters used are similar to that of Fig.\ref{fig2}.}
\label{fig3}
\end{figure*}

Fig.~\ref{fig3}(a) depicts $I^f(S;Y)$ and $I^f(X;Y)$ and both of them are $\approx 1$ for an extended regime of $\langle Y_s \rangle^f$ that takes into account active participation of both direct and indirect branches within C1 FFL. TSC and BN provide lesser values of these MI fractions. This gives a quantitative understanding of the network structure and function where two-variable MI becomes sufficient to replace the corresponding three-variable total MI. The parameters for which both $\langle Y_s \rangle$ and $\langle Y_x \rangle$ become significant, lead to form a C1 FFL motif with moderately strong direct and indirect pathways. The portion of the information content of S, which is filtered out depending upon the separation of relaxation time scales between S and X, now gets an alternative route to bypass X and directly feed gene Y. Our present construct sets $\mu_x = 5 \mu_s$ whereas $\mu_y = 50 \mu_s$ which enables Y to sample S much better than what X does. The FFL region showing $I^f(S;Y) \approx 1$ signifies $I(S;Y) \approx I(S;X,Y)$ according to Eq.(\ref{eq7a}). The relevance of X is consequently diminished as uncertainty reduction of S using X becomes minimal. This gets reflected in $I^f(S;X|Y) \approx 0$ where the fraction is taken with respect to I(S;X,Y) (see Fig.\ref{fig3}(b)). Since I(S;X,Y) is finite, this means $I(S;X|Y) \approx 0$ (see Eq.(\ref{eq6a})) indicative of an approximate Markov chain $S \rightarrowtail Y \rightarrowtail X$ i.e., S and X are conditionally independent of each other provided Y is known \cite{Cover1991}. Interestingly enough, this parametric domain also contributes maximum $I^f(X;Y)~(\approx 1)$ implying $I(X;Y) \approx I(X;Y,S)$ following Eq.(\ref{eq7b}). In this FFL region, I$^f$(X;Y) practically overlaps with I$^f$(S;Y). Correspondingly in Fig.\ref{fig3}(b), we note that $I(X;S|Y) \approx 0$, i.e., there exists another approximate Markov chain $X \rightarrowtail Y \rightarrowtail S$. To understand the significance of the principle MI fractions i.e., $I^f(S;Y)$ and $I^f(X;Y)$, we draw attention to the Venn diagrams presented in Figs.(\ref{figa1}-\ref{figa2}). Taking into account Eq.(\ref{eq6a}) along with Fig.\ref{figa1}(a) and fig.\ref{figa2}(a), it is evident that $I(S;Y) \leq I(S;X,Y)$. Similarly, Eq.(\ref{eq6b}) in conjunction with Fig.\ref{figa1}(b) and Fig.\ref{figa2}(b) clearly dictates that $I(X;Y) \leq I(X;Y,S)$. Near perfect equalities in both of these cases are achieved only for C1 FFL, indicating that both direct and indirect decoding of the signal is equally efficient in this motif. In contrast, TSC (BN) possesses no direct (indirect) decoding machinery.

\subsection{Output noise: effect of architectural modifications on the target gene expression}

To support our information-theoretic findings showing enhanced information transduction in C1 FFL, we define Output Noise expressed in the units of the inverse of copy number (c.n.): 
\begin{equation}  
Output~Noise =: \Sigma(Y)/(\langle Y_s \rangle \langle Y_x \rangle).
\end{equation} 
Output Noise quantifies changes in the output gene-expression fluctuations resulting from master (co) regulator enhancing (diminishing) their individual contributions in synthesizing copies of gene-product Y i.e., $\langle Y_s \rangle~(\langle Y_x \rangle)$. We present profile of Output Noise responding to ${\langle Y_s \rangle}^f$ in Fig.\ref{fig3}(c). The two extreme points, TSC ($\langle Y_s \rangle=0$) and BN ($\langle Y_x \rangle=0$) force Output Noise to blow up at ${\langle Y_s \rangle}^f=0~and~1$, respectively (data points are not shown). As may be inferred from Fig.\ref{fig2}(c), ${\Sigma(Y)}_{OSC}/(\langle Y_s \rangle \langle Y_x \rangle)~({\Sigma(Y)}_{TSC}/(\langle Y_s \rangle \langle Y_x \rangle))$ will have a steeper ascend (descend) with increasing ${\langle Y_s \rangle}^f$ while ${\Sigma(Y)}_{CT}/(\langle Y_s \rangle \langle Y_x \rangle)$ will remain constant (see Eqs.(\ref{eq5a}-\ref{eq5d})). The former two terms dominating over the third in the two extreme points ultimately provide the concave down profile of Output Noise showing that the noise floor (minimal noise) belongs to C1 FFL while TSC and BN provide high noise walls. This signature of lessening in output fluctuations is positively correlated with increased information transmission capacity (higher $I^f(S;Y)$ and $I^f(X;Y)$) of C1 FFL. Likewise noisy TSC and BN harness lesser amounts of these MI fractions. 


\begin{figure*}[!t]
\includegraphics[width=1.8\columnwidth,angle=0]{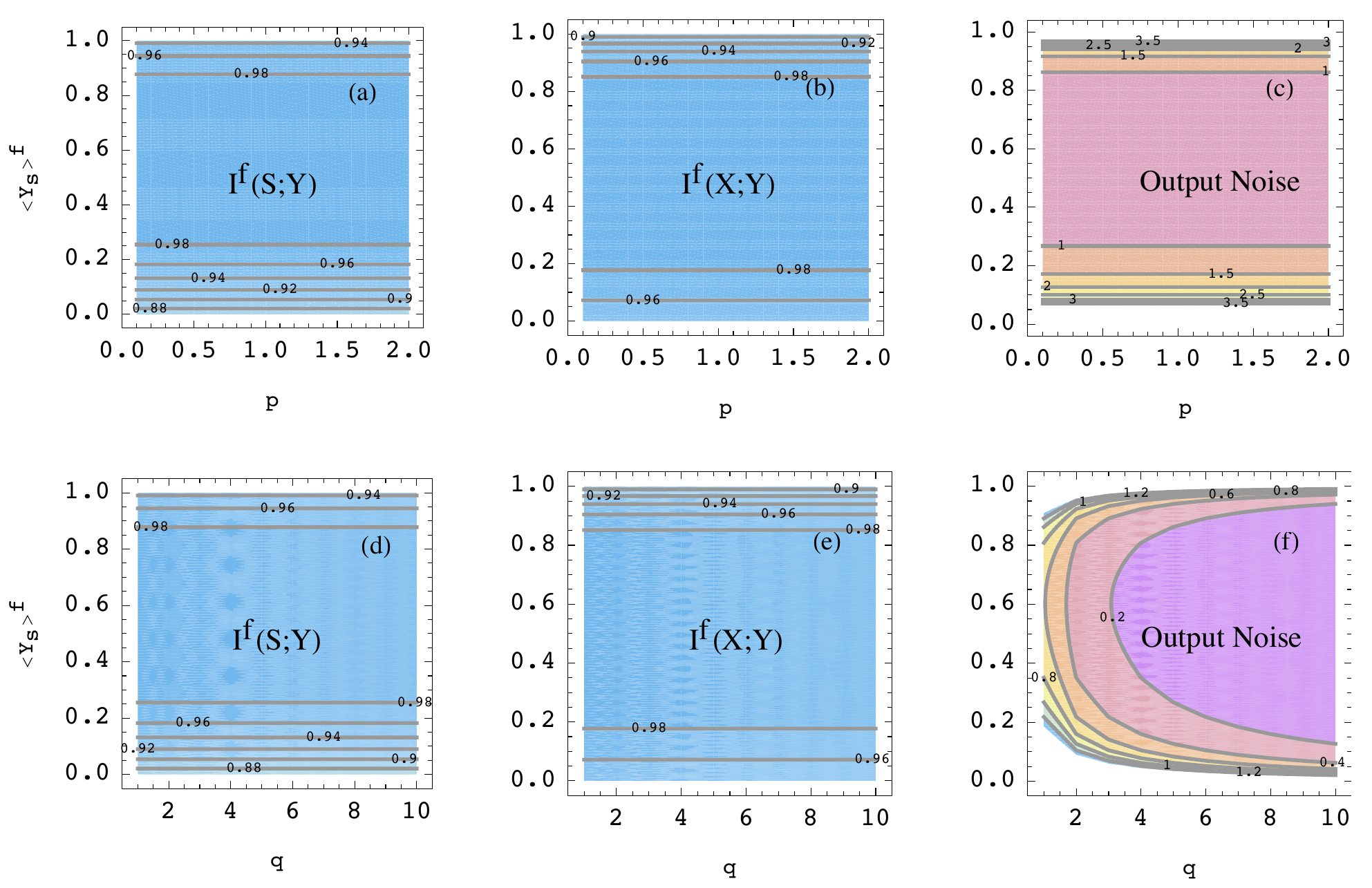}
\caption{Parametric dependence of the results obtained in Fig.\ref{fig3}(a) and Fig.\ref{fig3}(c). Panels (a)-(c): $I^f(S;Y)$, $I^f(X;Y)$, and Output Noise, respectively as functions of ${\langle Y_s \rangle}^f$ and `p' for the transformation $\mu_i \rightarrow p \mu_i$, where $i \in \{s,x,y\}$. Population levels, activation coefficients, and `n' are same as in Figs.(\ref{fig2},\ref{fig3}). Panels (d)-(f): behavior of the previous metrics with respect to ${\langle Y_s \rangle}^f$ and `q' for the transformation $\langle Z \rangle \rightarrow q \langle Z \rangle$, where $Z \in \{S,X,Y\}$. Relaxation rate parameters, dependence of activation coefficients on population levels, and `n' are kept identical with Figs.~(\ref{fig2},\ref{fig3}).
\label{fig4}
}
\end{figure*}

\subsection{Parametric dependence of MI fractions and Output Noise}

Next, to investigate the parametric dependence of our core results, depicted in Fig.\ref{fig3}(a,c), we focus on relaxation rate parameters and steady-state population levels because in our analysis these two types of entities are preset independently and the synthesis rate parameters are determined as a consequence of fixed steady-state gene-expression criteria. Firstly, we undertake a transformation $\mu_i \rightarrow p \mu_i~\forall ~i \in\{s,x,y\}~and~p\in {\mathbb R}^{+}$. Under this condition, all relaxation rate parameters are made `p' times of their values in our original computation (as in Figs.(\ref{fig2}-\ref{fig3})), keeping all the population levels fixed at values as in the previous profiles, we quantify the responses of $I^f(S;Y)$, $I^f(X;Y)$, and Output Noise with respect to ${\langle Y_s \rangle}^f$ in Figs.\ref{fig4}(a-c). When compared with their counterparts in Fig.\ref{fig3}, the contour levels in these plots running parallel to the `p' axis, reveal that our previous findings in Fig.\ref{fig3} are invariant of identical scaling in $\mu_i$. For analytical support, one can take a look at Eqs.(\ref{eq3a}-\ref{eq3f}), consisting of different $\Phi_{ij}$ which remain unchanged under the current type of transformation for all the $\mu_i$. With increasing `p', all the gene-products are becoming less stable, keeping their relative separation of time scales fixed with the same magnitude of filtering of upstream signal all over the `p' range.

Taking to the next check-point of parametric dependence, we introduce another transformation of the type $\langle Z \rangle \rightarrow q \langle Z \rangle~\forall~Z \in\{S,X,Y\}~and~q\in {\mathbb Z}^{+}$, while all $\mu_i$ are fixed at values as in Fig.\ref{fig3}. In other words, all steady-state population levels are made `q' times of the previous values used in our preceding analysis of Figs.(\ref{fig2}-\ref{fig3}). $I^f(S;Y)$ and $I^f(X;Y)$ in Figs.\ref{fig4}(d,e), respectively, show identical behavior compared to their counterparts in Fig.\ref{fig3}. Here also, all the contours are manifested parallel to `q' axis attesting invariance of these MI fractions with respect to the current transformation. Although this invariance is absent from Output Noise profile in Fig.\ref{fig4}(f), we have similar concave down characteristics of the plot as in Fig.\ref{fig3}(c). We notice that with increasing `q' i.e., increasing population levels of all the gene-products by the same factor lowers the noise floor pertaining to C1 FFL. Taking recourse to Eqs.(\ref{eq3a}-\ref{eq3f}) and a bit of further straight-forward algebra, it can be shown that unlike the $\mu_i$ transformation case, the second moments do not remain unchanged under $\langle Z \rangle$ transformation. This `q' dependence only exits at the level of computing MI, each of which involves ratios of a variance of a species and a corresponding partial variance. Consequently, $I^f(S;Y)$ and $I^f(X;Y)$ keep identical functional dependence on ${\langle Y_s \rangle}^f$. With raising `q', increasing values of $\langle Z \rangle$ diminishes each of the terms of Output Noise summing in the reduced noise floor. In Eqs.(\ref{eq5b}-\ref{eq5d}), after dividing ${\Sigma(Y)_{OSC}}$, ${\Sigma(Y)_{TSC}}$, and ${\Sigma(Y)_{CT}}$ by $\langle Y_s \rangle\langle Y_x \rangle$ (for brevity, let us call these rescaled quantities as ${\Sigma(Y)^{'}_{OSC}}$, ${\Sigma(Y)^{'}_{TSC}}$, and ${\Sigma(Y)^{'}_{CT}}$, respectively), we observe their contributions in different regions of Fig.\ref{fig4}(f), designated by the values of `q' and ${\langle Y_s \rangle}^f$. In upper (lower) most part of Fig.\ref{fig4}(f), ${\Sigma(Y)}^{'}_{OSC}~({\Sigma(Y)}^{'}_{TSC})$ dominates over ${\Sigma(Y)}^{'}_{TSC}~({\Sigma(Y)}^{'}_{OSC})$. The variable parts in ${\Sigma(Y)}^{'}_{OSC}$ are ${\langle Y_x \rangle}^{-1}$ and $\langle Y_s \rangle{\langle Y_x \rangle}^{-1}{\langle S \rangle}^{-1}$ whereas in ${\Sigma(Y)}^{'}_{TSC}$ those are ${\langle Y_s \rangle}^{-1}$, $\langle Y_x \rangle{\langle Y_s \rangle}^{-1}{\langle X \rangle}^{-1}$, and $\langle Y_x \rangle{\langle Y_s \rangle}^{-1}{\langle S \rangle}^{-1}$. On the other hand, ${\Sigma(Y)}^{'}_{CT}$ involves ${\langle S \rangle}^{-1}$ and as a result of which contributes steadily across the entire range of ${\langle Y_s \rangle}^{f}$ for a fixed `q'. These observations are drawn consulting Eqs.(\ref{eq5b}-\ref{eq5d}). Now in the upper most part of Fig.\ref{fig4}(f) (${\langle Y_s \rangle}^{f}$ maximum), as `q' changes from 1 to 10, $\langle Y_s \rangle$ changes from 99 to 999 ($\langle Y_x \rangle$ remains fixed at 1) which along with a 10 times increase in $\langle S \rangle$ contribute little towards the dominating ${\Sigma(Y)}^{'}_{OSC}$. Similarly in the lower most part of Fig.\ref{fig4}(f) (${\langle Y_s \rangle}^{f}$ minimum), $\langle Y_x \rangle$ increases from 99 to 999 ($\langle Y_s \rangle$ remains fixed at 1). This along with 10 times increase in $\langle S \rangle$ and $\langle X \rangle$ only give a nudge to the primary contributor ${\Sigma(Y)}^{'}_{TSC}$. It is observed that as one approaches intermediate zone of ${\langle Y_s \rangle}^{f}$, the contours become more distinguished from each other. In and around ${\langle Y_s \rangle}^f=$ 0.5, considering q : $1\rightarrow10$, ${\Sigma(Y)}^{'}_{OSC}$ and ${\Sigma(Y)}^{'}_{CT}$ get affected by a 10 times increase in $\langle S \rangle$ while ${\Sigma(Y)}^{'}_{TSC}$ involves 10 times increase in both $\langle S \rangle$ and $\langle X \rangle$. Besides, ${\Sigma(Y)}^{'}_{OSC}$ and ${\Sigma(Y)}^{'}_{TSC}$ also involve 10 times increase in $\langle Y_x \rangle$ and $\langle Y_s \rangle$, respectively. Taking into consideration all these changes in the perspective of the variable factors in ${\Sigma(Y)}^{'}_{OSC}$, ${\Sigma(Y)}^{'}_{TSC}$, and ${\Sigma(Y)}^{'}_{CT}$, the Output Noise profile in Fig.~\ref{fig4}(f) may be justified. Figs.~\ref{fig4}(d,e) dictate that by equal fold-change of biochemical populations involved, the signal decoding abilities quantified by MI fractions can not be increased. On the other hand, keeping in mind that the essential source of biological noise is actually the low copy numbers of different biochemical species, this behavior of Output Noise as in Fig.\ref{fig4}(f) can be readily justified. The conditional MI fractions being related to the above-mentioned MI fractions through Eqs.(\ref{eq6a}-\ref{eq6b}), the invariant features of major MI fractions also hold true for the conditional ones.


\begin{figure*}[!t]
\includegraphics[width=1.8\columnwidth,angle=0]{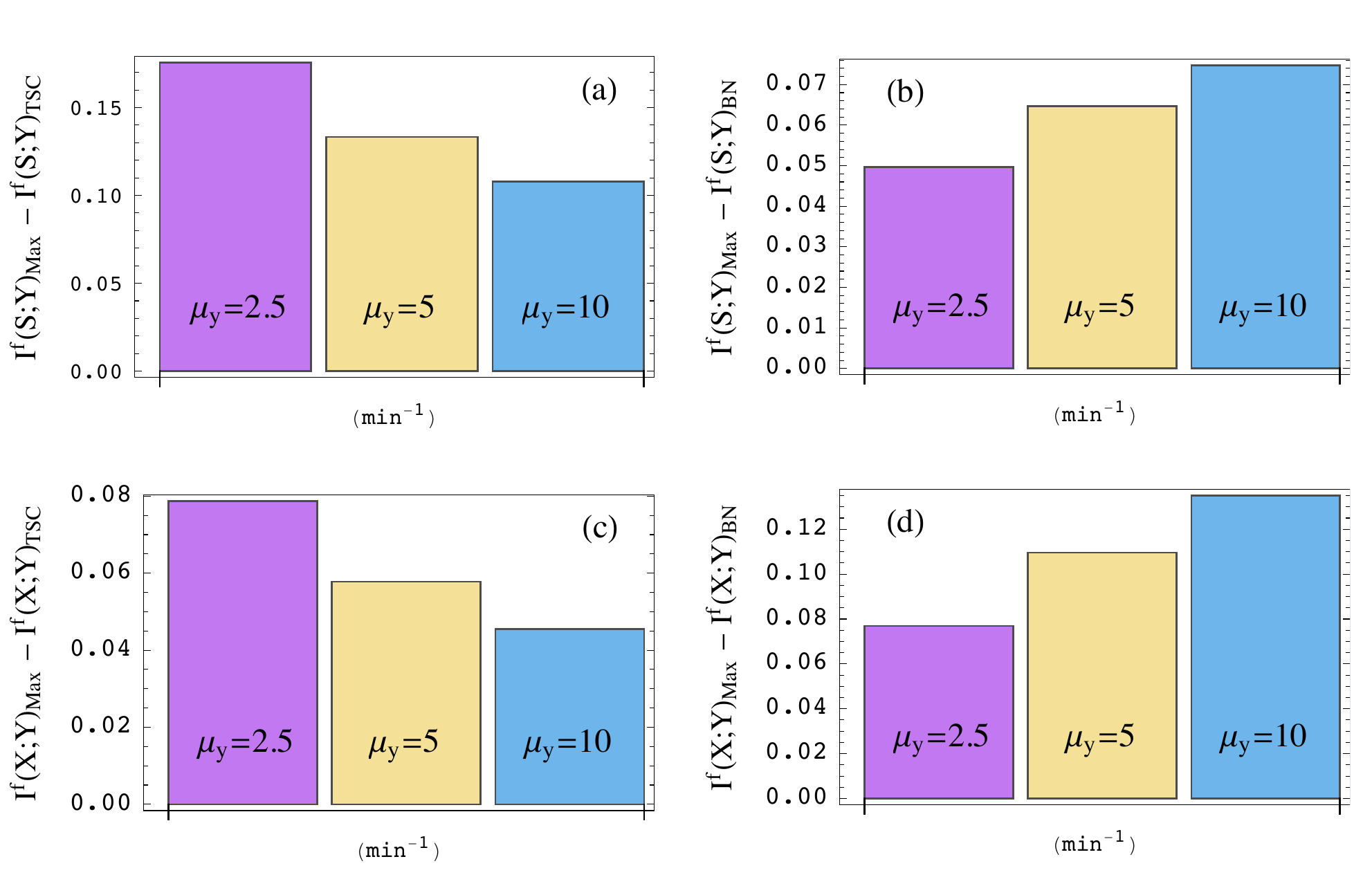}
\caption{These bar charts quantitate difference between MI fractions contributed by TSC (${\langle Y_s \rangle}^f=0$), BN (${\langle Y_s \rangle}^f=1$) and the corresponding maximum (denoted `Max') value of MI fractions coming from the FFL domain while ${\langle Y_s \rangle}^f$ is varied from 0 to 1. Three different values of the output relaxation rate parameter ($\mu_y=2.5,~5,~10~min^{-1}$) are considered to showcase that the gain in MI fractions considering variations in ${\langle Y_s \rangle}^f$ is controlled by the stability of the final gene-product. Other parametric constraints are identical with Figs~(\ref{fig2}-\ref{fig3}).  
\label{fig5}
}
\end{figure*}

Fig.~\ref{fig5} demonstrates the parametric dependence of the gain in MI fractions as the three-node network pattern transforms itself starting from TSC and finally becoming BN, passing through C1 FFL. The gain is calculated using the maximum MI fractions coming from the FFL region and each of the corresponding contributions from TSC (${\langle Y_s \rangle}^f=0$) and BN (${\langle Y_s \rangle}^f=1$). The control parameter here is $\mu_y$, i.e., the relaxation rate parameter of the output gene-product. It is observed that, as $\mu_y$ increases, the gain with respect to TSC is diminished both for $I^f(S;Y)$ and $I^f(X;Y)$ as depicted in Fig.~\ref{fig5}(a,c). Fig.\ref{fig5}(b,d) documents the increase in MI fractions with respect to BN, and both have increasing trends responding to increase in $\mu_y$. It is to be noted that for the two new cases of $\mu_y=2.5~and~5~min^{-1}$, we have checked that they produce similar profiles of statistical metrics as in Figs.~(\ref{fig2}-\ref{fig3}) (data is not shown here). These data sets point to the fact that whereas better information decoding capacity of C1 FFL with respect to both TSC and BN holds for a range of $\mu_y$ values, the relative performance of the former compared to the rest depends on the stability of the output gene-product.


\section{Conclusion}

In this article, we presented an information-theoretic framework which allows distinguishing a C1 FFL from a TSC and BN in the perspective of stochastic signal processing. Our findings show that neither two nor three-variable MI, but a fraction of the former to the latter has a central position in this regard. The set of biologically plausible parameters in our model provides maximal MI fractions between the target gene-product and each of the master and co-regulator of the output gene, for an extended region where both the direct and indirect branches of the FFL are reasonably well functional in synthesizing the target gene-product. This indicates in the ability of a C1 FFL to achieve near-perfect direct and indirect decoding of the environmental fluctuations encoded in the upstream gene-expression profiles. This has been supported by additional analysis concentrating upon the Output Noise level, which shows antagonistic response compared with that of the MI fractions. Our modeling approach has some unique parametric invariance properties which are reflected in the variational nature of the MI fractions and Output Noise. Interestingly enough, our predictions hold their grounds under specific transformations on relaxation rate parameters and gene-expression levels of all the constituent nodes in the present generic network which gives rise to TSC, C1 FFL, and BN. These signatures of invariance are supposed to be biologically meaningful as the former rate parameter quantifies the stability of gene-products, whereas their expression levels are related to the environmental impact over the organismal fitness.

The present construct uses an additive signal integration mechanism in the target gene because it enables us to utilize all of the physiologically meaningful values in the entire range of ${\langle Y_s \rangle}^f$. This modeling flexibility is hindered in multiplicative signal integration which is also employed in gene-transcriptional network motifs. The restriction in the multiplicative case arises mainly because there both the master and co-regulator TFs are needed at the same time to produce the output gene-product. The present approach deals with single time-points at steady state and may be tested with extension involving multiple time-points and dynamic-states. Temporal structure in the gene-expression profiles may be better handled using the concept of information transfer which is connected with the causal aspect of signal processing than mere correlation which is better done with MI as in the present case. As the dynamical signals have been observed to yield higher information transmission \cite{Selimkhanov2014}, it may also affect the difference among MI fractions arising from TSC, C1 FFL, and BN. In this regard, there may be other contributing factors like taking into account information processing by the single cells in a population rather than measuring the population-averaged information to avoid underestimation \cite{Chevalier2015}. Our theoretical predictions may be tested in synthetic bacterial circuitries. \textit{In vivo}, C1 FFL may be embedded in larger networks involving feedback and auto-regulation, and it would be interesting to check whether our present framework remains tenable, taking into account these additional interactions.

\begin{acknowledgments}
Md SAM is supported by DST, Govt. of India, through INSPIRE research fellowship (DST/INSPIRE Fellowship/2018/IF180056). AB and SKB acknowledge Bose Institute, Kolkata for research support.
\end{acknowledgments}


\appendix*
\renewcommand{\thefigure}{A\arabic{figure}}
\setcounter{figure}{0} 

\section{Venn diagrammatic representation of various MI}
To elucidate the concepts of two and three-variable MI as described in the main text and used in our present analysis, we present a number of Venn diagrams \cite{Cover1991,MacKay2002} as follows:

\begin{figure}[!h]
\subfigure[]{\includegraphics[scale=0.4]{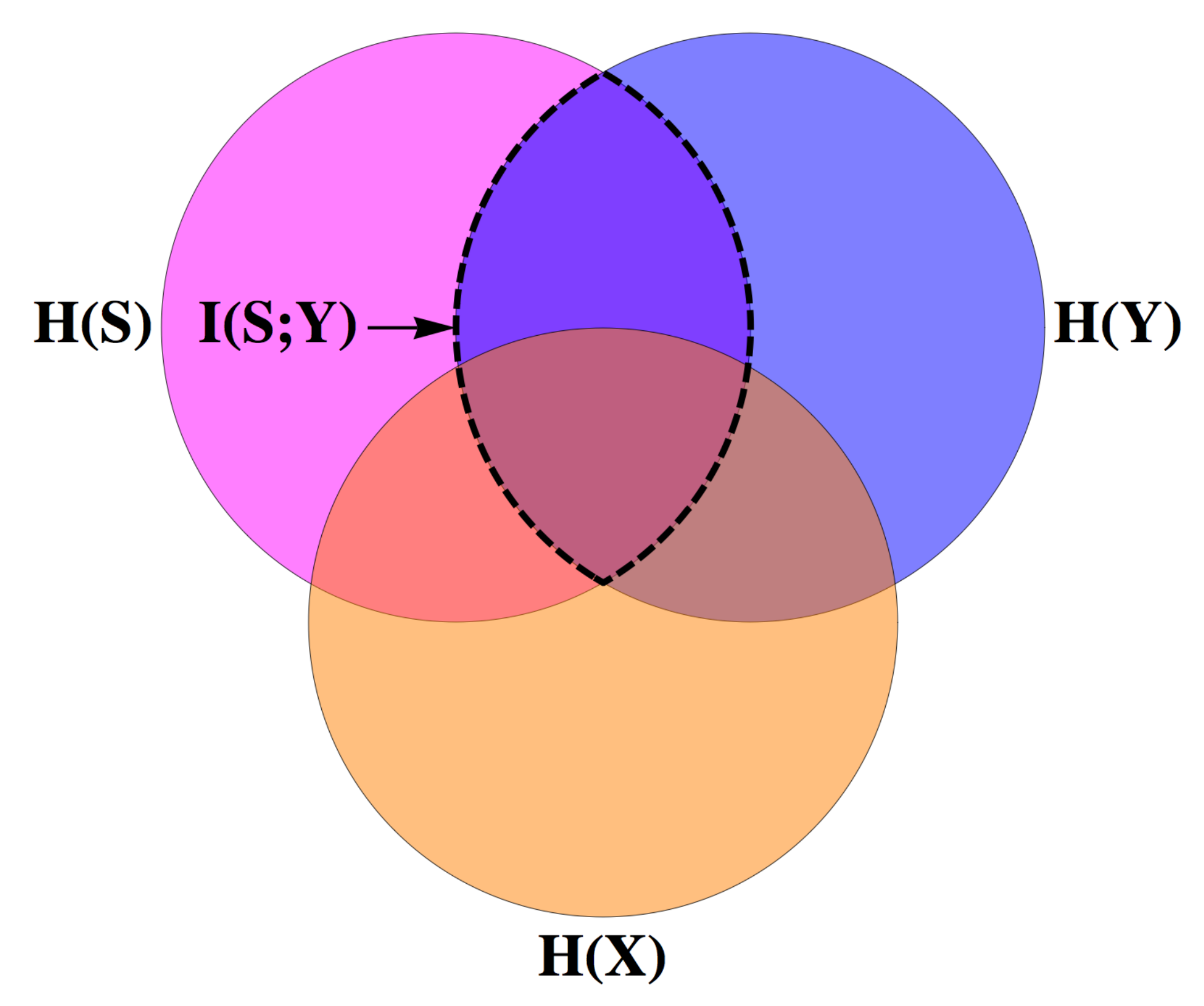}}
\subfigure[]{\includegraphics[scale=0.4]{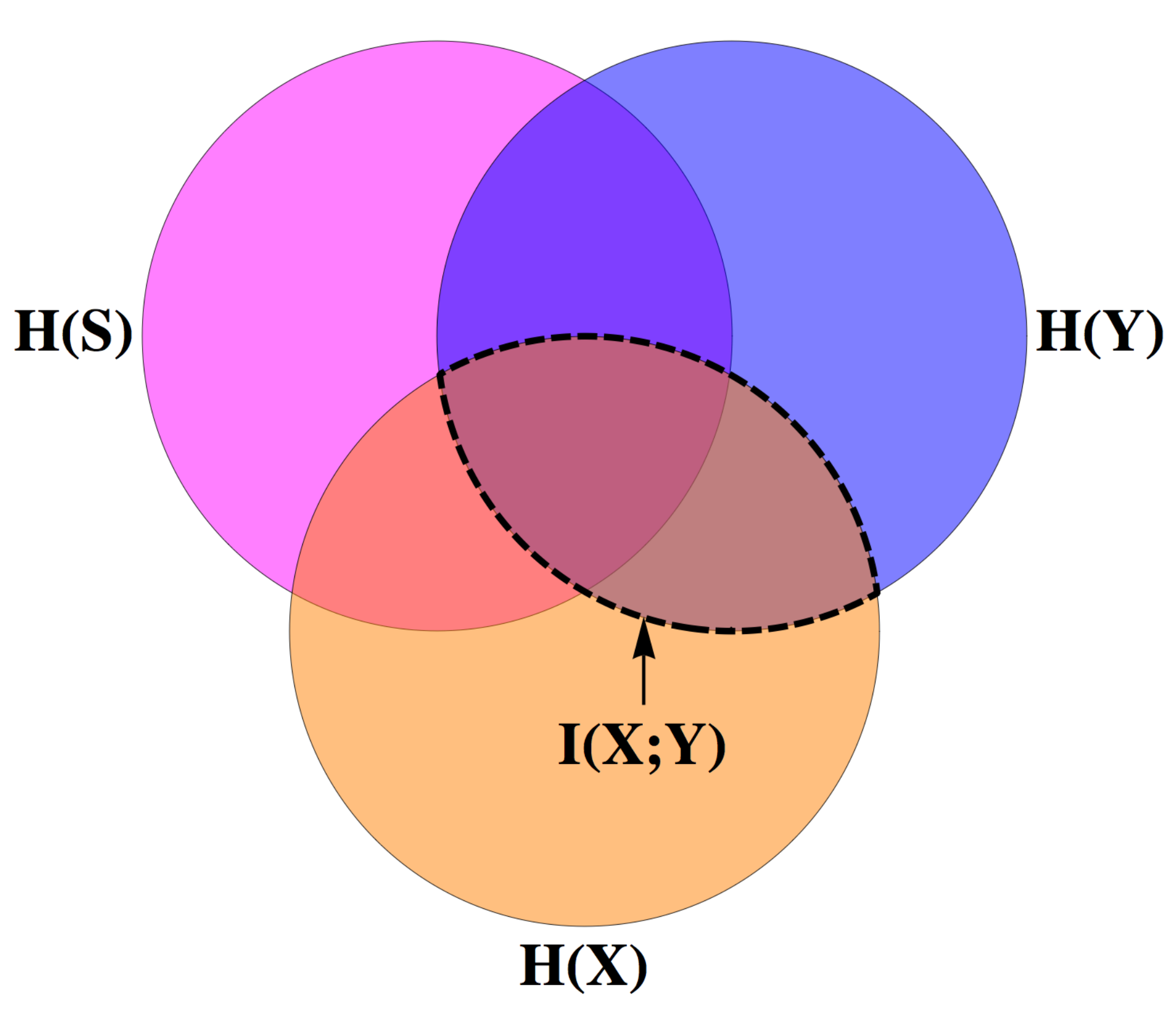}}
\caption{(color online) (a) The overlap between entropy spaces of S and Y i.e., H(S) and H(Y), respectively, denotes MI of S and Y i.e., I(S;Y). (b) Overlap region of H(X) and H(Y) denotes I(X;Y). Both MI spaces are enclosed by thick dashes.}
\label{figa1}
\end{figure}

\begin{figure}[!h]
\subfigure[]{\includegraphics[scale=0.4]{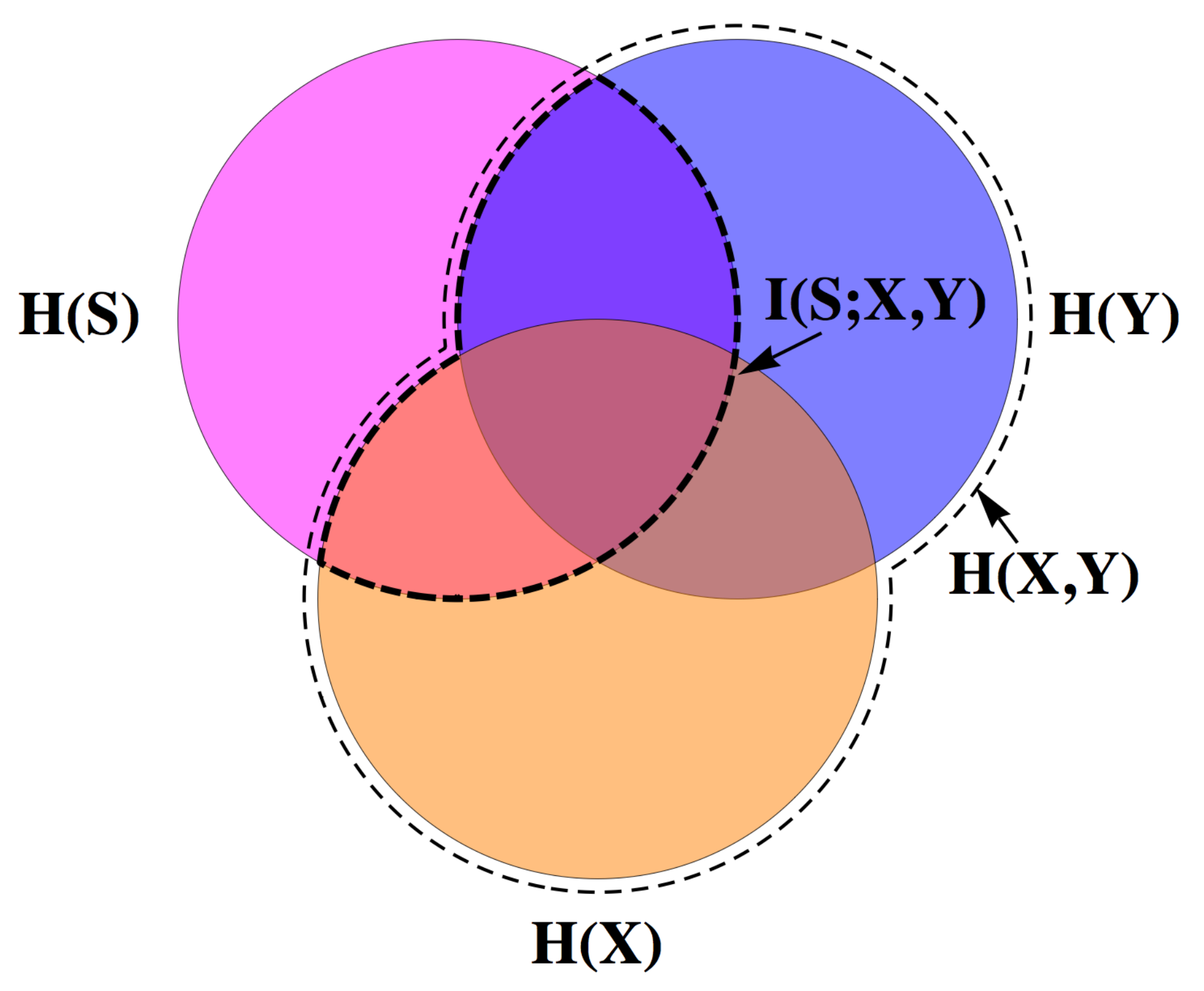}}
\subfigure[]{\includegraphics[scale=0.4]{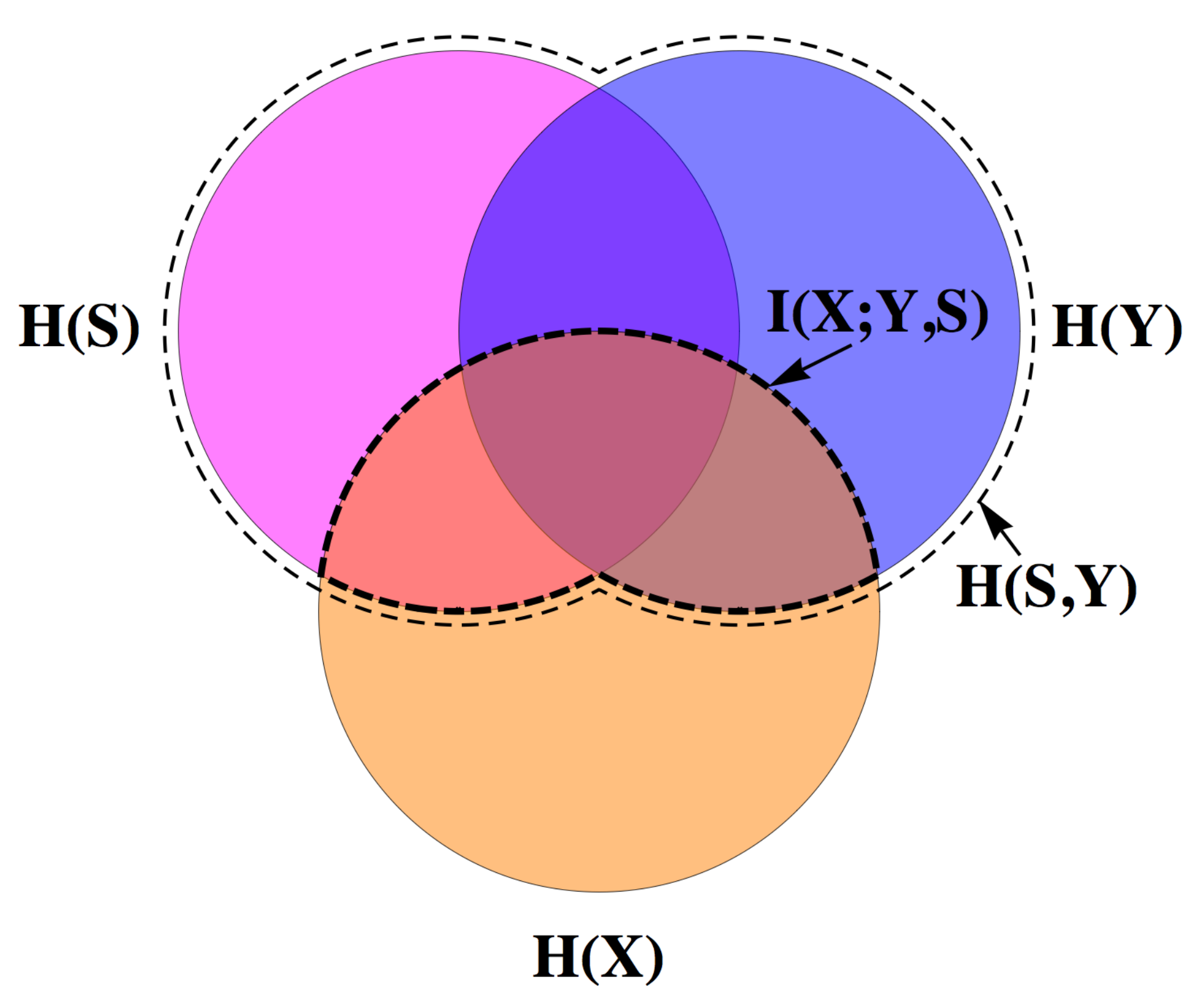}}
\caption{(color online) (a) The common region between H(S) and H(X,Y) (region enclosed by thin dashes) is the MI of S and (X,Y) i.e., I(S;X,Y). Similarly in panel (b), the common space between H(X) and H(S,Y) (region enclosed by thin dashes) signifies I(X;Y,S). Both three-variable MI regions are enclosed by thick dashes.}
\label{figa2}
\end{figure}

\begin{figure}[!h]
\subfigure[]{\includegraphics[scale=0.4]{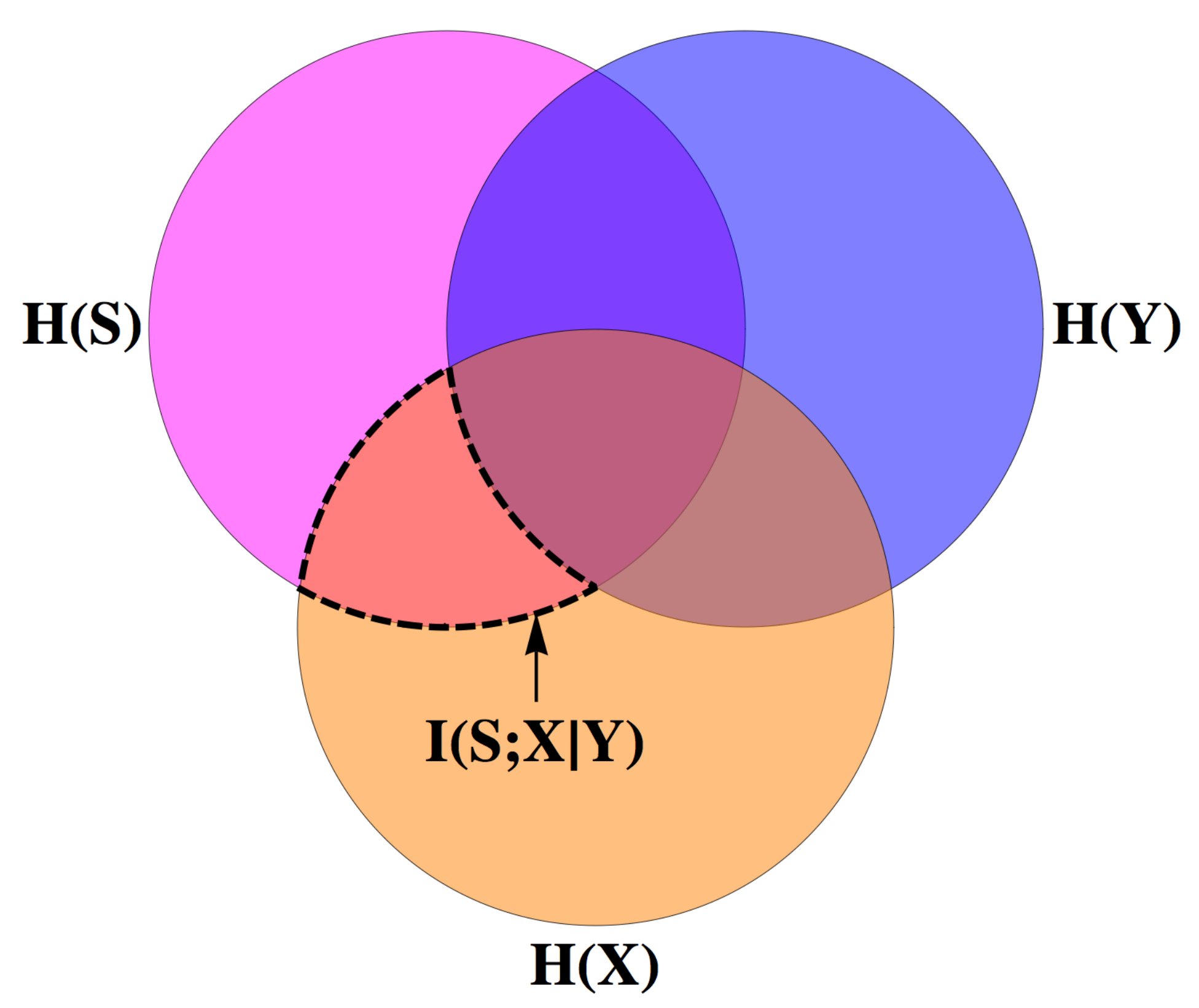}}
\caption{(color online) Region marked by thick-dashed enclosure depicts MI I(S;X$|$Y) which signifies shared space between H(S) and H(X) when Y is known. Comparing I(S;X,Y) in Fig.~\ref{figa2}(a) and I(S;Y) in Fig.~\ref{figa1}(a), it is easily observed that $\text{I(S;X$|$Y)} = \text{I(S;X,Y)} - \text{I(S;Y)}$. It is to be noted that $\text{I(X;S$|$Y)} = \text{I(S;X$|$Y)}$, verifiable by subtracting I(X;Y) in Fig.~\ref{figa1}(b) from I(X;Y,S) in Fig.~\ref{figa2}(b).}  
\label{figa3}
\end{figure}

\clearpage


\end{document}